\documentclass[aps,prd,nofootinbib]{revtex4}
\usepackage{graphicx}
\usepackage{amsfonts}
\usepackage{amsmath}
\usepackage{amssymb}

\begin{document}

\title{Thin accretion disks around a Schwarzschild acoustic black hole}
\author{R.M. Yusupova}
\email{yu.rose@mail.ru}
\affiliation{Zel'dovich International Center for Astrophysics, Bashkir State Pedagogical University, 3A, October Revolution Street, Ufa 450077, RB, Russia}
\author{R.N. Izmailov}
\email{izmailov.ramil@gmail.com}
\affiliation{Zel'dovich International Center for Astrophysics, Bashkir State Pedagogical University, 3A, October Revolution Street, Ufa 450077, RB, Russia}
\author{R.Kh. Karimov}
\email{karimov_ramis_92@mail.ru}
\affiliation{Zel'dovich International Center for Astrophysics, Bashkir State Pedagogical University, 3A, October Revolution Street, Ufa 450077, RB, Russia}

\date{18 June 2026}

\begin{abstract}
We investigate the properties of a thin accretion disk in the Novikov-Thorne framework around a Schwarzschild acoustic black hole characterized by a nonzero background fluid four-velocity $v_{r} = \sqrt{2M\xi/r}$, where $\xi$ is a tuning parameter. The Novikov-Thorne formalism is used here phenomenologically, with the Schwarzschild acoustic black hole treated as an effective background geometry for comparing disk observables with those of the Schwarzschild black hole. We show that the acoustic black hole exhibits two regimes depending on $\xi$. For $0 < \xi \leq 2.8$, the disk is in a strengthening regime, in which the radiative characteristics are enhanced relative to the Schwarzschild case. For $\xi > 2.8$, the disk is in a weakening regime, in which the radiative characteristics are reduced relative to the Schwarzschild case. The most pronounced change in the radiative characteristics is associated with the behaviour of the marginally stable orbit radius $r_{ms}$, whose branch change near $\xi\approx 2.8$ produces a sharp modification of the flux, temperature, and luminosity profiles. For the numerical illustrations, we adopt a toy model with black hole mass $M = 15M_{\odot}$ and compare the corresponding flux, temperature, spectral luminosity, efficiency, and Eddington luminosity with those of the Schwarzschild black hole. These differences may provide phenomenological diagnostics for comparing effective compact object geometries.
\end{abstract}


\maketitle


\section{Introduction}
\label{sec1}
Black holes are among the most remarkable predictions of general relativity, although some of their key quantum effects, such as Hawking radiation, have not yet been directly detected despite compelling indirect evidence for black holes \cite{Akiyama:2019a, Akiyama:2019b, Akiyama:2022a}. An important advance was made by W.G. Unruh, who showed that the propagation of phonons in a moving fluid is mathematically equivalent to that of a massless scalar field in curved spacetime \cite{Unruh:1981, Jacquet:2020}. This insight established the framework of analogue gravity and introduced the concept of an acoustic black hole, in which a sonic horizon is formed when the fluid flow becomes supersonic \cite{Visser:1998, Visser:1999, Barceli:2019, Pandey:2025, Ge:2019, Yu:2019, Mukherjee:2025, Das:2004}. Since then, numerous theoretical and experimental studies have investigated analogue horizons and Hawking-like phenomena in optical, hydrodynamic, Bose-Einstein condensate, polaritonic, and related systems \cite{Philbin:2008, Rousseaux:2008, Faccio:2010, Lahav:2010, Parvizi:2023, Weinfurtner:2011, Nguyen:2015, Jiazhong:2019}.

A solution describing an acoustic black hole in Schwarzschild spacetime was proposed by Ge et al. \cite{Ge:2019}, namely, the Schwarzschild acoustic black hole (SABH) characterized by a nonzero background fluid four-velocity $v_{r} = \sqrt{2M\xi/r}$, where $\xi$ is a tuning parameter. This solution is derived from two fundamental frameworks: the Gross-Pitaevskii equation and a ($3+1$)-dimensional Einstein-Yang-Mills theory with an anti-de Sitter geometry \cite{Gross:1961}. Furthermore, subsequent work demonstrated that starting from the ($3+1$)-dimensional Yang-Mills theory, an effective ($2+1$)-dimensional acoustic metric can be obtained, which has potential applications within the holographic principle \cite{Ge:2019}.

The effective metric describes the background geometry on which fields propagate. In many analogue-gravity models, the effective metric is a kinematic property of the background flow. While the metric is often derived by linearizing the massless Klein-Gordon equation, the resulting geometry $g_{\mu\nu}$ can serve as a fixed background for other fields. In this work, the SABH metric is treated as an effective test geometry. We do not claim a literal astrophysical accretion disk made of phonons or other analogue-medium excitations. Rather, we ask how the standard Novikov-Thorne disk observables would change if equatorial circular motion were governed by the SABH line element instead of the Schwarzschild one. The SABH should be understood as an effective analogue-gravity configurations, in which a suitable fluid velocity profile generates acoustic perturbations obeying a wave equation formally analogous to that in curved spacetime, rather than as genuine gravitational black holes.

Treating the SABH as an effective geometry, one may ask how standard thin disk observables differ from those of the Schwarzschild case. As is well known, accretion is an efficient energy release mechanism in stellar systems, influencing the evolution of binary stars, galactic nuclei, and the formation of jets. Various models describe these accretion processes, including the spherically-symmetric Bondi-Hoyle model \cite{Bondi:1952, Michel:1972, Bahamonde:2015, Yusupova:2021, Yusupova:2025}, ADAF (Advection-Dominated Accretion Flow) models \cite{Narayan:1994, Narayan:1995, Narayan:2008}, and thin \cite{Shakura:1973, Novikov:1973, Page:1974, Thorne:1974}, thick \cite{Wang:2025b, Chen:2024, Wei:2023}, and general relativistic magnetohydrodynamic (GRMHD) disks \cite{Jiang:2019, Dhruv:2025, Nampalliwar:2022}.

Despite the current prevalence of sophisticated numerical simulations, finding analytical solutions remains crucial for understanding the fundamental physics of accretion. The pioneering studies of accretion disks using a Newtonian approach were conducted by Shakura and Sunyaev \cite{Shakura:1973}. Subsequently, a fully general relativistic model for thin disks was developed in the seminal works of Novikov and Thorne \cite{Novikov:1973}, Page and Thorne \cite{Page:1974}, and Thorne \cite{Thorne:1974}. This model assumes a stationary state with a constant mass accretion rate independent of the disk radius. We shall focus on it in this paper to see how accretion profiles of the SABH differ from those of Schwarzschild black hole (hereinafter SBH). Applications of the Novikov-Thorne accretion model are well established in the literature, e.g., the study of properties of accretion disks around black holes \cite{Kumar:2025, Bogush:2022, Heydari:2024, Heydari:2023, Boshkayev:2024, Alloqulov:2024, Mirzaev:2023, Wu:2024, Li:2025a, Feng:2025, Cui:2026, Liu:2025}, exotic central objects such as wormholes \cite{Harko:2008, Karimov:2020, Karimov:2022, Rahaman:2021}, gravastars and naked singularities \cite{Harko:2009b, Torres:2002, Kurmanov:2025}, magnetically charged black holes in string theory in \cite{Karimov:2018}. Recent extensions of the Novikov-Thorne thin disk formalism to non-Kerr and modified-gravity backgrounds include hybrid metric-Palatini black holes \cite{Dyadina:2024}, Hartle-Thorne spacetimes \cite{Boshkayev:2024}, regular black holes sourced by nonlinear electrodynamics \cite{Kurmanov:2024, Khoshrangbaf:2025, Uniyal:2023, Cai:2025}, and rotating hairy Horndeski black holes \cite{Heydari:2024, Li:2025b}, where deviations in ISCO location, radiative flux, spectral luminosity, and efficiency have been systematically explored.

The objective of this work is to investigate how standard Novikov-Thorne thin disk observables are modified when the Schwarzschild geometry is replaced by the SABH effective geometry. This classical model, based on a well-established system of equations, provides precise, testable predictions for the emission spectrum and radial temperature profile of an accretion disk. By applying this model, we aim to systematically investigate and quantitatively assess the influence of the tuning parameter on the flow kinematics and the resulting emission spectra of the SABH. Previous research has established key features of SABHs. It was shown in \cite{Vieira:2021, Vieira:2022} that the signal of the quasinormal-mode signal of the SABH is weaker than that of Schwarzschild black hole. Toshmatov et al. \cite{Toshmatov:2023} extended the perturbative spectroscopy of SABH to massive scalar and electromagnetic fields, finding isospectral electromagnetic parity sectors, longer-lived massive scalar quasi-resonances, and linear stability under the perturbations studied. They scan the acoustic tuning parameter mainly over $\xi \in [0,4]$, with the $\xi < 2$ regime singled out as reducing both the oscillation frequency and damping rate of the electromagnetic/scalar perturbations. The work provides a useful theoretical benchmark for analogue gravity ringdown studies. Gravitational lensing of SABH was investigated in \cite{Qiaoa:2023}. Qiao showed that when tuning parameter satisfies $0 < \xi < 3.32$, there is only one circular photon orbit, which is unstable, and when $\xi > 3.32$, the acoustic Schwarzschild black hole may have three circular photon orbits (two unstable ones and a stable one in between). Ditta et al. \cite{Ditta:2023} study particle dynamics and weak gravitational lensing in an SABH surrounded by plasma, analyzing effective potentials and ISCO behavior for massive particles and photons. Galactic microlensing by SABH was studied in \cite{Akhtaryanova:2026}. The Bondi accretion of different kinds of fluid (ultra-stiff fluid, ultra-relativistic fluid, radiation fluid, and sub-relativistic fluid) was studied in \cite{Mukherjee:2025} around the acoustic charged black hole. By varying the tuning parameter $\xi$, Mukherjee et al. obtained stable results that are compatible with the observational data. Toshmatov and Abdukarimov \cite{Toshmatov:2022} investigated massive and massless particle motion and photon frequency shifts from an accretion disc, mainly for $0\le\xi\le3$. Ditta et al. further analysed particle motion, ISCO properties, and plasma lensing in the acoustic Schwarzschild geometry.

In this paper we study the kinematic and emissivity properties of a central object represented by SABH using the Novikov-Thorne model. We consider the parameter range $0\le\xi\le6$, identify the transition in the selected marginally stable orbit branch near $\xi\simeq2.8$, and compute the main characteristic of the disk: flux, temperature, spectral luminosity, accretion efficiency and Eddington luminosity. The principal new physical result is the separation of the SABH disk diagnostics into a strengthening regime for $0<\xi\lesssim2.8$ and a weakening regime for $\xi\gtrsim2.8$. In particular, we wish to see how the kinematic and emissivity properties compare with the Schwarzschild case and identify the transition associated with the change of the selected $r_{ms}$ branch. We shall assume for numerical illustration a toy model of a central object with mass $15M_{\odot}$ and accretion rate $\dot{M}_{0}=10^{19}\,\mathrm{g\,s^{-1}}$.

The novelty of this paper lies in applying the Novikov-Thorne thin disk formalism to the Schwarzschild acoustic black hole metric, viewed as an effective background geometry. Since the acoustic metric is derived from massless perturbations, while thin accretion disks are composed of massive matter, this application is necessarily phenomenological rather than microscopic. Nevertheless, near the black-hole horizon the motion of infalling particles may become ultrarelativistic, with kinetic energy greatly exceeding rest mass energy \cite{Bisnovatyi:2011}. In such a regime, the dynamics can be expected to depend more strongly on the background geometry and less strongly on the particle rest mass. Motivated by this observation, we investigate how the standard disk characteristics such as the marginally stable orbit, flux, temperature, luminosity, and efficiency are modified when the Schwarzschild geometry is replaced by its acoustic counterpart.

We emphasize the restricted interpretation adopted in this work. The acoustic metric is a kinematical effective metric for perturbations of a background medium. Therefore, ordinary baryonic matter in an astrophysical spacetime would not, without an additional coupling mechanism, be forced to follow geodesics of the acoustic metric. In the present paper we use the SABH line element as an effective test geometry and assume that the disk degrees of freedom entering the Novikov-Thorne equations are effective test matter minimally coupled to $g_{\mu\nu}^{\rm SABH}$. With this assumption, the standard circular geodesic quantities and the flux formula can be evaluated formally. The resulting flux, temperature, luminosity, and efficiency should therefore be read as phenomenological metric diagnostics, not as a claim that a laboratory acoustic flow contains an ordinary massive astrophysical accretion disk.

The paper is organized as follows. In Sec. 2, we briefly review the SABH solution. In Sec. 3, we summarize the main formulas of the Novikov-Thorne model used in this work. In Sec. 4, we discuss the Eddington luminosity. In Sec. 5, we summarize the results. We take units such that $G = c = 1$, metric signature ($-,+,+,+$) and greek indices run from $0$ to $3$.

\section{Acoustic black hole solution}
\label{sec2}

The SABH metric is given by \cite{Ge:2019, Qiaoa:2023}
\begin{eqnarray}
ds^{2} &=& -f(r)dt^{2} + f^{-1}(r)dr^{2} + r^{2}\left( d\theta^{2} + \sin^{2}\theta d\varphi ^{2}\right) , \\
f(r) &=&\left( 1-\frac{2M}{r}\right) \left[ 1-\xi \frac{2M}{r}\left( 1 - \frac{2M}{r}\right) \right],
\end{eqnarray}%
where $M$ is the mass, $\xi$ is the tuning parameter. It is obvious that when $\xi =0$, the metric in Eqs. (1)-(2) reduces to the SBH metric. When $\xi\rightarrow +\infty$, the whole spacetime ($0\leq r\leq +\infty $) is inside the acoustic black hole.

In the equatorial plane, the metric components of (1)-(2) reduce, respectively, to
$$g_{tt}=-\left( 1-\frac{2M}{r}\right) \left[ 1-\xi \frac{2M}{r}\left( 1-\frac{2M}{r}\right) \right],$$
$$g_{rr}=\left[\left( 1-\frac{2M}{r}\right) \left\{ 1-\xi \frac{2M}{r}\left( 1-\frac{2M}{r}\right) \right\}\right]^{-1},$$
$$g_{\theta \theta }=r^{2},$$
$$g_{\varphi \varphi }=r^{2}\sin ^{2}\theta.$$

The event horizon is a null surface determined by the equation $f(r)=0$. The equation has three roots: optical horizon $r_{s}$, interior $r_{\textmd{ac}_{-}}$ and exterior horizons $r_{\textmd{ac}_{+}}$
\begin{equation}
r_{s}=2M,
\end{equation}
\begin{equation}
r_{\textmd{ac}_{\pm}} = M \left( \xi \pm \sqrt{\xi ^{2}-4\xi }\right) .
\end{equation}

A necessary condition for the existence of an acoustic event horizon is $\xi \geq 4$. When the tuning parameter satisfies $0\leq \xi <4$, only the "optical" horizon exists in SABH, but both of the interior $r_{\textmd{ac}_{-}}$ and exterior $r_{\textmd{ac}_{+}}$ horizons disappear. When $\xi =4$, the interior and exterior event horizons coincide with each other, and we get the extreme SABH. When $\xi >4$, there are three regions in SABH. In region $r<r_{s}$, both light rays (photons) and sound waves (phonons) cannot escape from the SABH. In region $r_{s} < r < r_{\textmd{ac}_{+}}$, light rays could escape from the SABH, while sound waves cannot. In region $r > r_{\textmd{ac}_{+}}$, both light rays and sound waves could escape from the SABH \cite{Qiaoa:2023}.

\section{Thin accretion disk: Novikov-Thorne model}
\label{sec3}

\subsection{Thin disk approximations}
\label{sec3-1}

The Novikov-Thorne model describes geometrically thin, optically thick, cold accretion disks around compact objects. The basic equations of the Novikov-Thorne model, comprising the conservation laws, equation of state, and radiation laws, rely on the following assumptions \cite{Page:1974, Faraji:2020}:

1. The disk height $H$ above the equator is much smaller than the characteristic radius $R$ of the disk, $H\ll R$;

2. The disk is formed by particles moving in nearly Keplerian orbits around the central compact object;

3. The disk is in a state of hydrostatic equilibrium, which stabilizes its vertical structure, with pressure and the vertical entropy gradient being negligible;

4. An efficient radiative cooling mechanism operates within the disk, where heat is lost through radiation from its surface. This process prevents the accumulation of heat generated by viscous stresses and dynamic friction;

5. The time-averaged rest-mass accretion rate $dM_{0}/dt$ is independent of the radius: $\dot{M_{0}} \equiv dM_{0}/dt = -2\pi ru^{r}\Sigma = \textmd{const}$ (here $t$ and $r$ are the coordinate time and radial coordinates respectively, $u^{r}$ is the radial component of the four-velocity $u^{\mu}$ of the accreting particles and $\Sigma $ is the averaged surface density of the disk);

6. The disk is optically thick in the vertical direction, meaning radiation is thermalized and the emission is close to a blackbody spectrum;

7. The mass of the disk itself is negligible compared to the mass of the central object, so the disk's own gravity does not influence the spacetime metric.

\subsection{Kinematic properties}
\label{sec3-2}

Under the aforementioned assumptions, we consider a static, spherically symmetric geometry with metric of the general form
\begin{equation}
ds^{2} = g_{tt}dt^{2} + g_{rr}dr^{2} + g_{\theta\theta} d\theta^{2} + g_{\varphi\varphi}d\varphi^{2}.
\end{equation}

At and around the equator, i.e., when $\left\vert \theta -\pi /2\right\vert \ll 1$, we assume, with Harko \textit{et al.} \cite{Harko:2008}, that the metric
functions $g_{tt},g_{rr},g_{\theta \theta }$ and $g_{\varphi \varphi }$
depend only on the radial coordinate $r$.

Integration of the geodesic equation yields the effective potential $V_{\textmd{eff}}(r)$, given by \cite{Torres:2002}
\begin{equation}
V_{\textmd{eff}}\left( r\right) = g_{tt}\left( 1+\frac{\widetilde{L}^{2}}{%
g_{\varphi \varphi }}\right),
\end{equation}%
where $\widetilde{L}$ is the specific angular momentum of particles moving in circular orbits.

Existence of circular orbits at any arbitrary radius $r$ in the equatorial plane demands that $V_{\textmd{eff}}\left( r\right) =\widetilde{E}^2$ and $dV_{\textmd{eff}}/dr=0$. These conditions allow us to write the angular velocity $\Omega$, the specific energy $\widetilde{E}$, and the specific angular momentum $\widetilde{L}$ of particles moving in
circular orbits for SABH 
\begin{eqnarray}
&&\Omega =\frac{d\varphi }{dt}=\pm \frac{\sqrt{-g_{tt,r}g_{\varphi \varphi
,r}}}{g_{\varphi \varphi ,r}}=\sqrt{\frac{M\left\{12M^{2}\xi +r(r-8M\xi +r\xi )\right\}}{%
r^{5}}}, \\ 
&&\widetilde{E}=-\frac{g_{tt}}{\sqrt{-g_{tt}-g_{\varphi \varphi }\Omega ^{2}}%
}=\frac{(r-2M)\left\{r^{2}-2M(r-2M)\xi \right\}}{r^{2}\sqrt{r^{2}+16M^{2}\xi - 20M^{3}\xi / r - 3Mr(1+\xi )}}, \\
&&\widetilde{L}=\frac{g_{\varphi \varphi }\Omega }{\sqrt{-g_{tt}-g_{\varphi
\varphi }\Omega ^{2}}}=\sqrt{\frac{Mr^{2}\left\{12M^{2}\xi -8Mr\xi +r^{2}(1+\xi )\right\}}{r^{3}-20M^{3}\xi +16M^{2}r\xi -3Mr^{2}(1+\xi )}}.
\end{eqnarray}

The behaviour of the potential $V_{\textmd{eff}} \left( r\right)$ in Eq. (6) determines the marginally stable radius $r_{\textmd{ms}}$, or innermost stable circular orbit (ISCO) as the solution of
\begin{equation}
\frac{d^{2}V_{\textmd{eff}}}{dr^{2}} = g_{tt, rr} + \widetilde{L}^{2} \left( g_{tt} g_{\varphi\varphi}^{  -1}\right)_{,rr} = 0,
\end{equation}%
while the orbits at higher radii are Keplerian. The kinematic parameters in (8) and (9) do not impose restrictions on the tuning parameter, i.e., $\xi$ can take both positive and negative values. In our paper, we will choose positive values of $\xi$, as in, for example, \cite{Qiaoa:2023}.
\begin{figure}[!ht]
  \centerline{\includegraphics[scale=0.35]{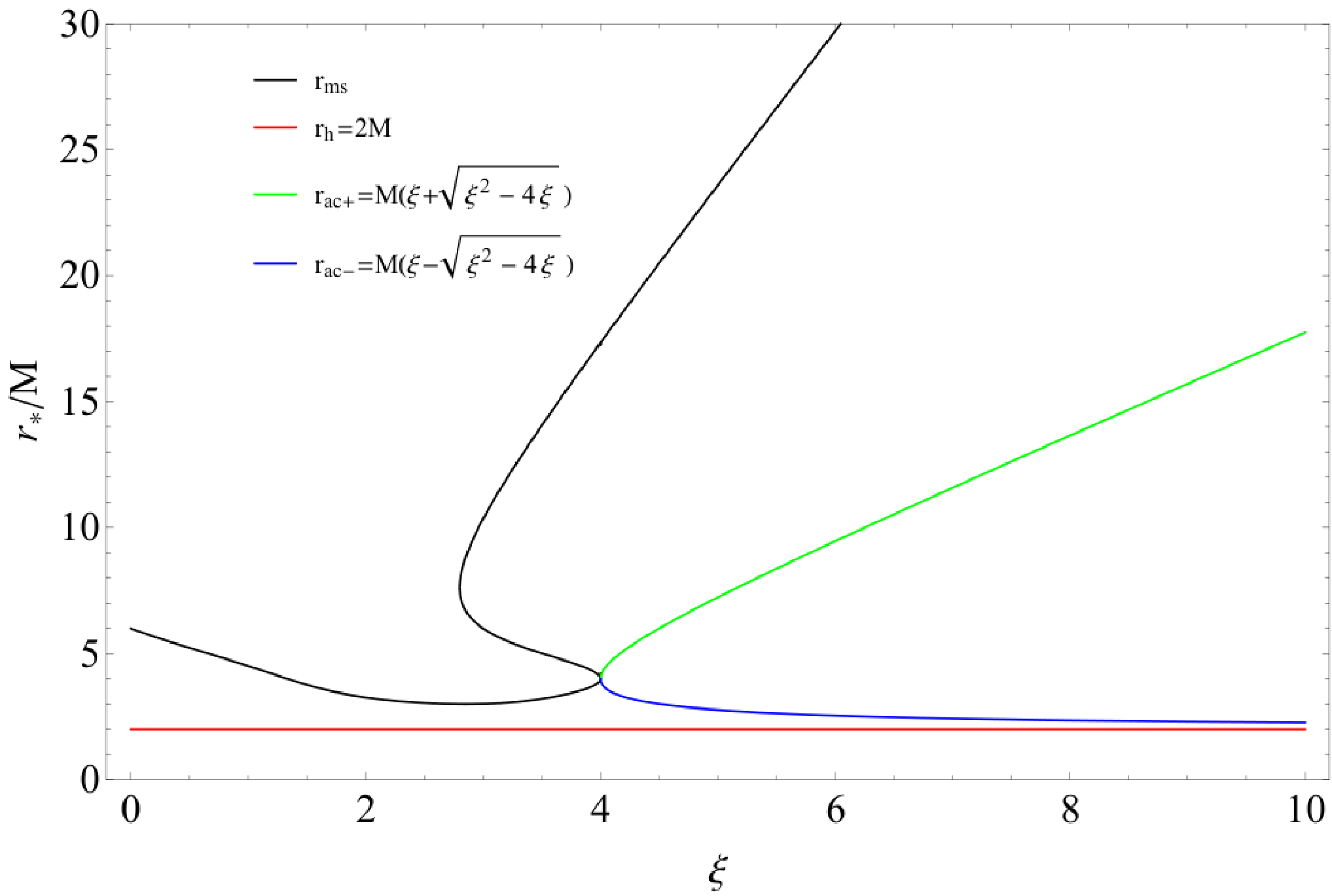}}
  \caption{The marginally stable orbit radius $r_{ms}$, the optical event horizon $r_{s}$, outer acoustic horizon $r_{\textmd{ac}_{+}}$ and inner acoustic horizon $r_{\textmd{ac}_{-}}$ as a function of the tuning parameter $\xi$.}
\end{figure}

Fig. 1 shows the dependence of the marginally stable orbit radius $r_{ms}$, the optical horizon $r_{s}$, and the acoustic horizons $r_{\textmd{ac}_{\pm}}$ on the tuning parameter $\xi$. As shown in Fig.1 at $\xi <2.8$, only an optical event horizon $r_{s}$ (red line) exists, along with a single possible marginally stable orbit radius $r_{ms}$ (black line). Furthermore, as the $\xi$ parameter increases, the value of $r_{ms}$ decreases. At $2.8 < \xi < 4$, there is also one optical event horizon $r_{s}$ but now three possible marginally stable orbit radii appear $r_{ms}$. At $\xi=4$, there is one event horizon $r_{s}$ and one acoustic horizon $r_{\textmd{ac}_{\pm}}$ (where the outer $r_{\textmd{ac}_{+}}$ (green line) and inner $r_{\textmd{ac}_{-}}$ (blue line) acoustic horizons coincide), as well as two possible radii of a marginally stable orbit $r_{ms}$. At $\xi >4$, all three horizons exist, but there is still only one possible marginally stable orbit radius $r_{ms}$. Furthermore, in this region, a linear dependence is observed, i.e., as the $\xi$ parameter increases, the value of the $r_{ms}$ also increases.

For $2.8<\xi<4$, the Eq.(10) admits three possible roots. Their physical role can be determined independently of the flux calculation by examining radial stability. In terms of $x = r/M$ and $\ell = L/M$, the condition $d\ell^2/dx>0$ gives stable circular orbits, while $d\ell^2/dx<0$ gives unstable circular orbits. The numerator of $d\ell^2/dx$ is $\mathcal{S}(x,\xi)$, given by
$$\mathcal{S}(x,\xi) = (1+\xi)x^5 -6(1+\xi)^2x^4 +(72\xi^2+60\xi)x^3 -(336\xi^2+80\xi)x^2 +672\xi^2x -480\xi^2.$$
The roots $x_1<x_2<x_3$ divide the circular orbit domain into an inner stable orbit, an intermediate unstable and an outer stable ones. Since the Novikov-Thorne disk is assumed to be supported from large radii and to remain continuous, its inner edge is the outermost marginally stable orbit $x_3$. The inner stable orbit is not used because it is disconnected from the outer disk by unstable circular orbits.

\begin{figure}[!ht]
  \centerline{\includegraphics[scale=0.37]{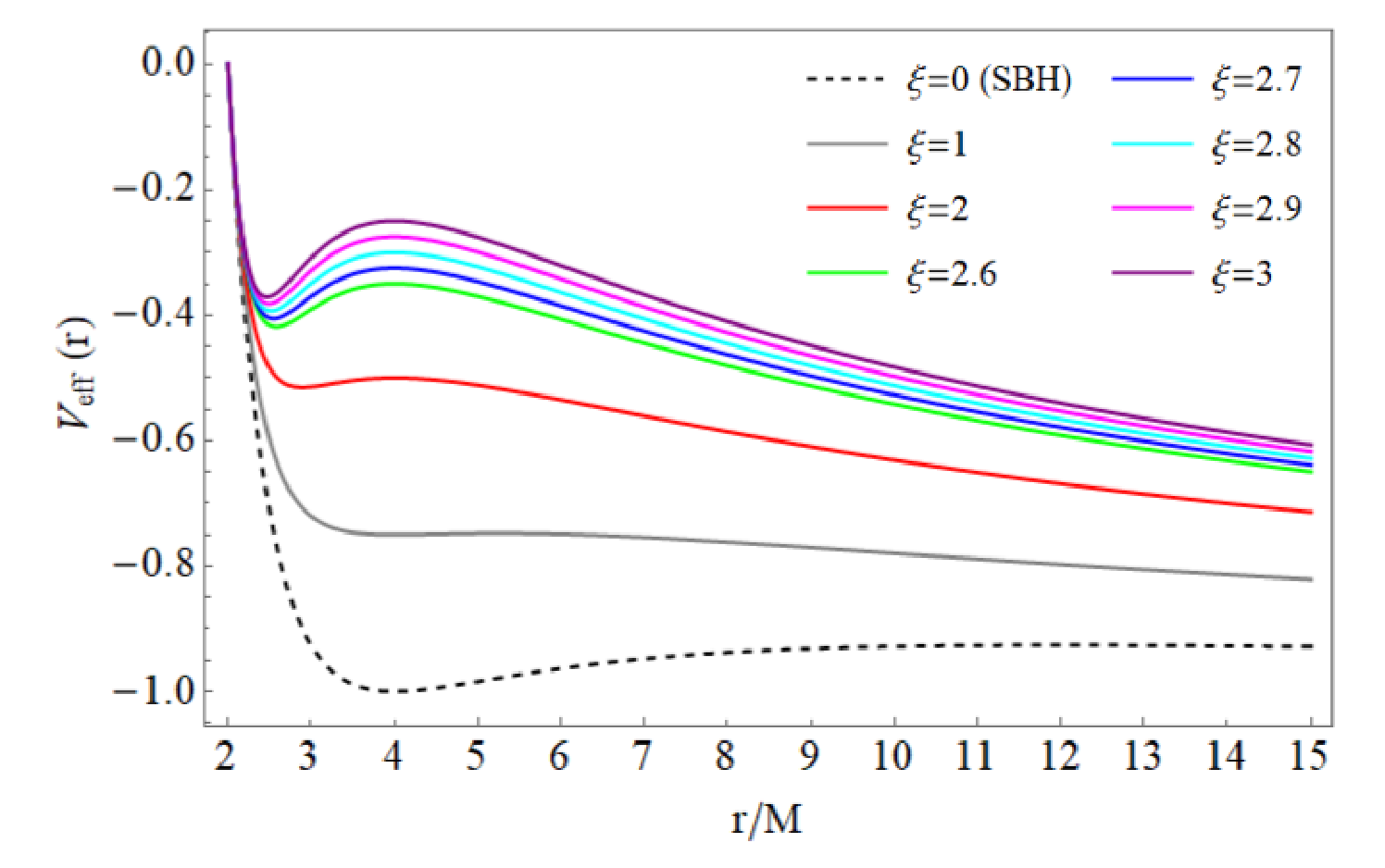} \includegraphics[scale=0.37]{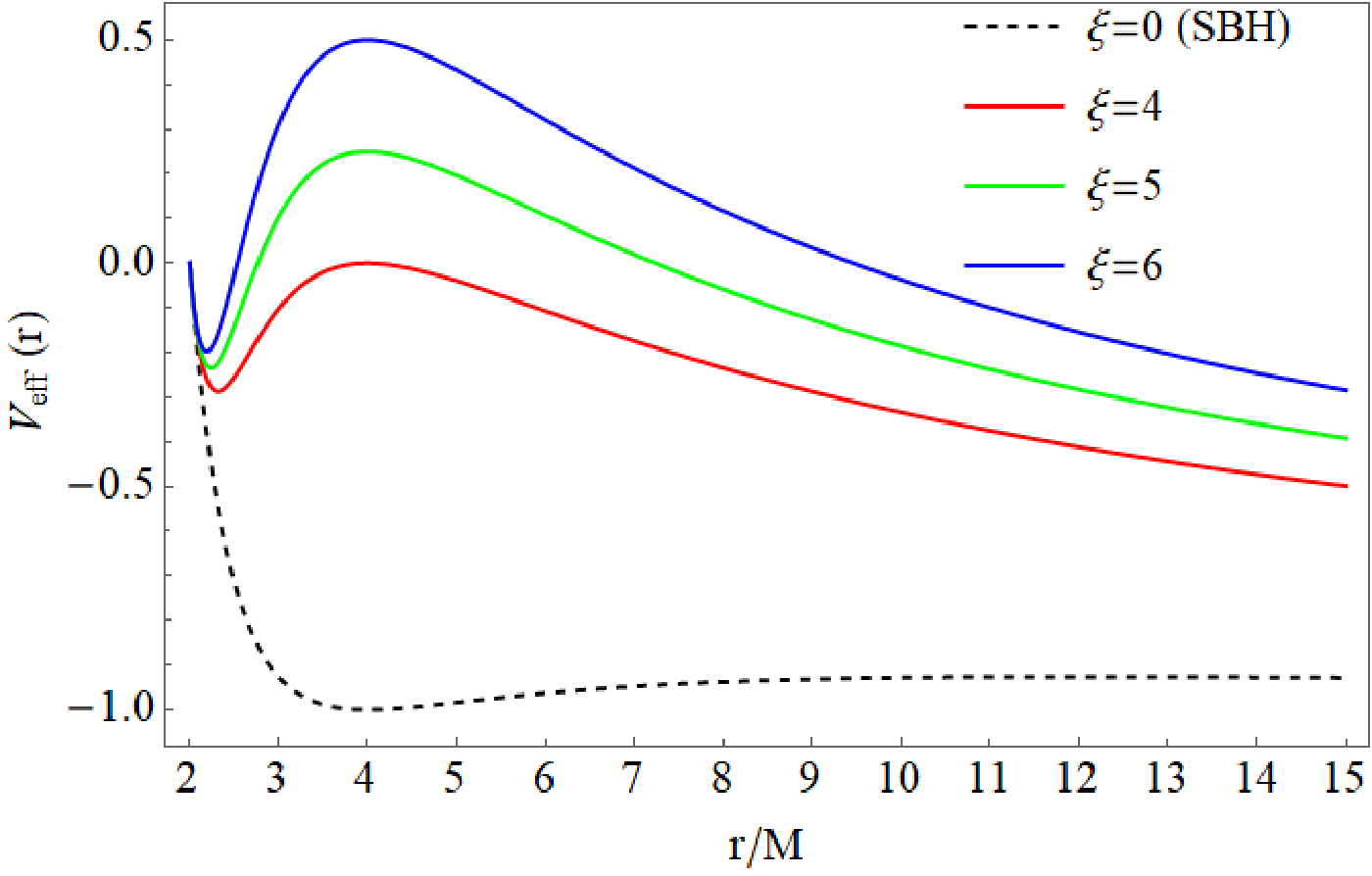}}
  \caption{The radial dependence of the effective potential of SABH for different values of the tuning parameter $\xi$. The specific angular momentum of the orbiting particle is chosen to be $\widetilde{L} = 4M$.}
\end{figure}

Next, we consider the radial dependence of the effective potential (6) of the SABH and compare it with the Schwarzschild black hole. The effective potential determines the geodesic motion of the test particles in the equatorial plane of the SABH. In the left plot of Fig. 2, we present the radial profile of the potential with different values of the tuning parameter $\xi<4$. The SABH, like the Schwarzschild black hole, has one minimum and one maximum in the effective potential. This means that both solutions have only one stable circular orbit. Note that the effective potential of the SABH is larger than that of the Schwarzschild black hole and an increase in $\xi$ leads to an increase in the effective potential. On the right plot of Fig. 2, we present the radial profile of the potential with different values of the tuning parameter $\xi \geq 4$. When $\xi = 4$, the maximum value of the effective potential is zero, as at the event horizon. Since in this case the maximum value coincides with the acoustic event horizon, then $V_{\textmd{eff}}|_{r=r_{h}} = V_{\textmd{eff}}|_{r=r_{_{\pm}}} = 0$. Note that when $\xi > 4$, the effective potential goes into the positive region in the interval $r\in(r_{\textmd{ac}_{-}}, r_{\textmd{ac}_{+}})$.

\subsection{Emissivity properties}
\label{sec3-3}

The emissivity properties of the accretion disk include: the flux $F(r)$, the temperature $T(r)$, and the luminosity $L\left( \nu \right)$. The flux $F(r)$ of the radiant energy
over the disk can be expressed in terms of $\Omega$, $\widetilde{E}$ and $\widetilde{L}$ as \cite{Harko:2008, Karimov:2022, Harko:2009b}
\begin{equation}
F\left( r\right) = - \frac{\dot{M_{0}}}{4\pi \sqrt{-g}}\frac{\Omega _{,r}}{\left( \widetilde{E}-\Omega \widetilde{L}\right)^{2}}\int_{r_{ms}}^{r} \left( \widetilde{E} - \Omega \widetilde{L}\right) \widetilde{L}_{,r}dr.
\end{equation}%
The disk is assumed to be in thermodynamic equilibrium, so the radiation flux emitted by the disk surface will follow the Stefan-Boltzmann law:
\begin{equation}
F\left( r\right) =\sigma T^{4}\left( r\right) ,
\end{equation}%
where $\sigma$ is the Stefan-Boltzmann constant, $T(r)$ is the temperature distribution.

The observed luminosity $L\left( \nu \right) $ has a redshifted black body spectrum \cite{Harko:2008}
\begin{equation}
L_{\nu }=4\pi d^{2}I(\nu )=\frac{8\pi h\cos {i}}{c^{2}} \int_{r_{\textmd{in}}}^{r_{\textmd{f}}}\int_{0}^{2\pi} \frac{\nu_{e}^{3}rdrd\varphi}{\exp\left[ \frac{h\nu _{e}}{k_{B}T}\right] -1},
\end{equation}%
where $i$ is the disk inclination angle, d is the distance between the observer and the center of the disk, $r_{\textmd{in}}$ and $r_{\textmd{f}}$ are the inner and outer radii of the disk, $h$ is the Planck constant, $\nu_{e}$ is the emission frequency, $I(\nu)$ is the Planck distribution, and $k_{B}$ is the Boltzmann constant. The observed photons are redshifted and their frequency $\nu$ is related to the emitted ones in the following way $\nu_{e} = (1+z)\nu$. The redshift factor $(1+z)$ has the form:
\begin{equation}
(1+z)=\frac{1+\Omega r\sin {\varphi }\sin {i}}{\sqrt{g_{tt}-\Omega
^{2}g_{\varphi \varphi }}},
\end{equation}%
where the light bending effect is neglected.

Another characteristic of the accretion disk is its efficiency $\epsilon$. The efficiency is measured at infinity and it is defined as the ratio of two
rates: the rate of energy of the photons emitted from the disk surface and the rate with which the mass-energy is transported to the central body. If
all photons reach infinity, an estimate of the efficiency is given by the
specific energy of the accreting particles measured at the marginally stable
orbit:
\begin{equation}
\epsilon =1-\widetilde{E}\left( r_{ms}\right) .
\end{equation}

The efficiency $\epsilon$ is an important characteristic, which quantifies the ability with which the central body converts the accreting mass into radiation.

Applying the above equations to the SABH spacetime, we obtain the corresponding emissivity profiles. Our results are described and plotted below. Note that in the range of $\xi$ from $2.8$ to $4$ there are three roots of $r_{ms}$ for a given value of $\xi$. Because a Novikov-Thorne disk supplied from large radii first encounters the outer stability boundary $x_3$, the disk inner edge is $x_3$. Flux positivity is then a consistency check. We also investigate the differences in the emissivity characteristics between the SABH and the SBH. We consider for illustration a central compact object of mass $M=15M_{\odot }$ with an accretion rate $\dot{M}_{0}=10^{19}\,\mathrm{g\,s^{-1}}$, and values of tuning parameter ($\xi =0,1,2,3,4,5,6$). The inner boundary is the outermost stable circular-orbit boundary outside the relevant horizon, for $\xi\geq4$, this requires $r_{\rm ms}>r_{\rm ac+}$, which, for the SABH solution depends on the tuning parameter $\xi $, as presented in Table 1. It is clear from the Table 1 that at $\xi \leq 2.8$ the efficiency $\epsilon$ of accretion as an energy release mechanism is strongly dependent on the compactness of the accreting object: the smaller the ratio $r_{ms}/M$, the greater the efficiency. At $\xi = 2.8$ there is a discontinuity in the function, since there is a jump in the choice of $r_{ms}$. At $\xi \geq 2.8$ the efficiency $\epsilon$ of accretion decreases as $r_{ms}$ increases. As $\xi$ tends to infinity, the efficiency of the SABH tends to the efficiency of the Schwarzschild black hole. Thus, the efficiency of the SABH is higher than that of a SBH for any value of $\xi$.

\begin{figure}[!ht]
  \centerline{\includegraphics[scale=0.18]{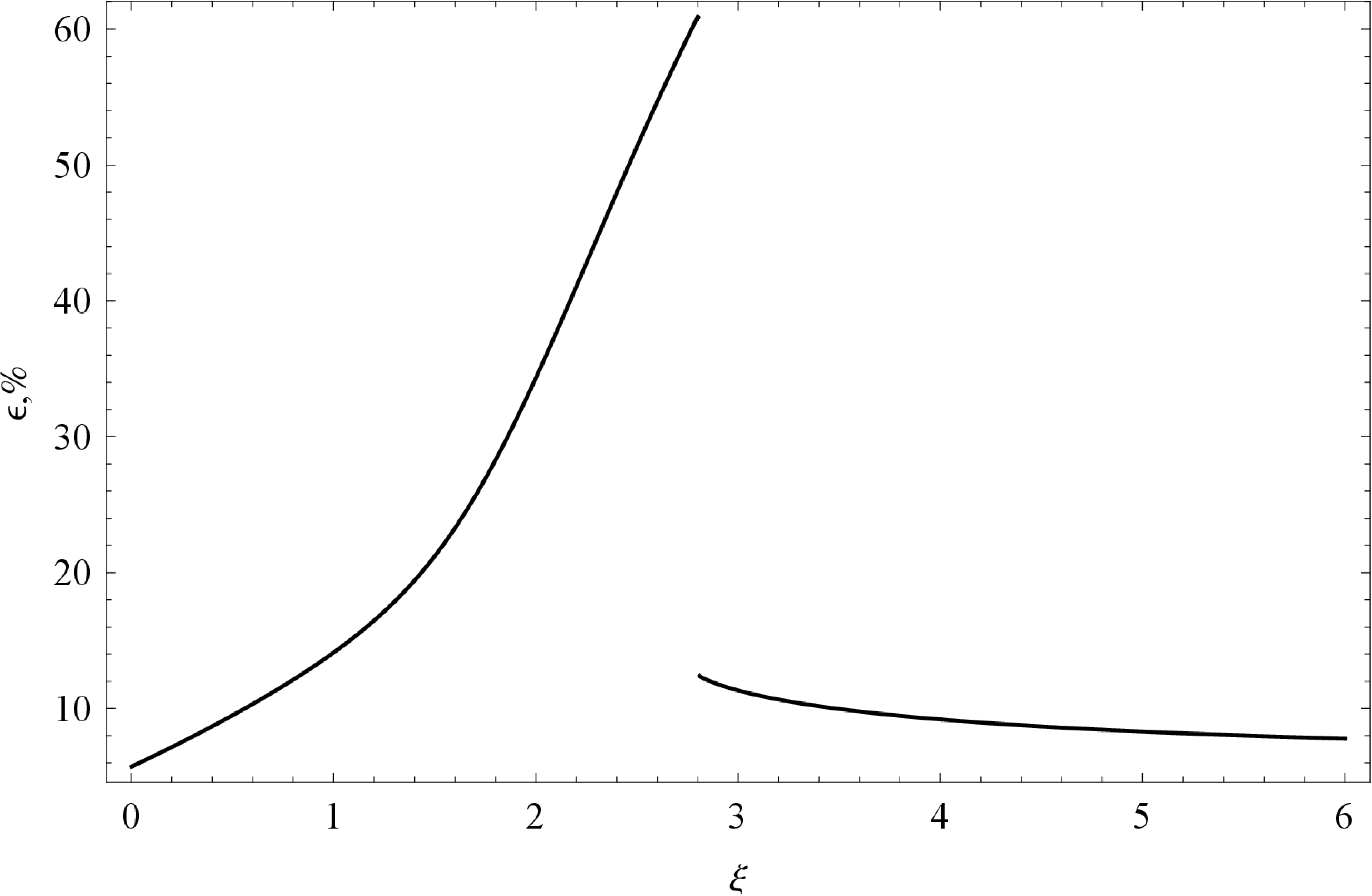}}
  \caption{The efficiency of SABH accretion disk.}
\end{figure}

\begin{table}[h]
\caption{The $r_{ms}$ and the efficiency $\epsilon $ for SABH ($\xi\neq 0$) and SBH ($\xi =0$).}\label{tab1}%
\centering
\begin{tabular}{@{}lll@{}}
\hline
$\xi$ & $r_{ms}/M$ & $\epsilon, \quad\%$ \\
\hline
$0$ (SBH) & $6$ & $5.719$ \\ 
$1$ & $4.5081$ & $14.116$ \\ 
$2$ & $3.2594$ & $34.331$ \\ 
$2.5$ & $3.03645$ & $51.357$ \\ 
$2.6$ & $3.0148$ & $54.653$ \\ 
$2.7$ & $3.000$ & $57.836$ \\ 
$2.8$ & $2.9922$ & $60.897$ \\ 
$2.81$ & $8.0459$ & $12.374$ \\ 
$2.82$ & $8.2882$ & $12.290$ \\ 
$2.85$ & $8.7799$ & $12.039$ \\ 
$2.9$ & $9.3811$ & $12.233$ \\ 
$3$ & $10.3379$ & $11.326$ \\ 
$4$ & $17.3354$ & $9.193$ \\ 
$5$ & $23.5934$ & $8.297$ \\ 
$6$ & $29.7137$ & $7.778$ \\
\hline
\end{tabular}
\end{table}

Fig. 4a shows the variation of time-averaged radiation flux $Log[F(r) / F_{\max ,SBH}(r)]$ with distance $Log[r/M]$ with increasing order of tuning parameter $\xi$. A generic qualitative feature of the plots with increasing values of $\xi $ to $2.8$ is that the profiles for SABH are always higher than those of the SBH and lower at $\xi > 2.8$. All the plots merge in the limit $r\rightarrow \infty$ since the SABH and SBH spacetimes are asymptotically flat. It is also important to note that as the $\xi$ increases, the maximum of the flux of radiation $F(r)$ shifts further from the center.\emph{\ }From the Table 2, we can see the exact ratios of the flux maxima $F_{\max}(r)$ as the tuning parameter $\xi$ increase. So, the ratios of fluxes $F_{\max}^{SABH}(r) / F_{\max}^{SBH}(r)\sim 4.83 \times 10^{3}$ at $\xi =2.8$ (SABH) and $\xi = 0$ (SBH), whereas $F_{\max}^{SBH}(r)/F_{\max }^{SABH}(r) = 91.2$ at $\xi = 6$ (SABH) and $\xi = 0$ (SBH). The exact quantitative ratios show how the radiation-flux profiles of the SABH can be lower than those of the SBH at values $\xi > 2.8$. This feature is consistently shared by other profiles too. A detailed analysis of the peak values shows that the radiation flux increases as $\xi$ increases from $0$ to $2.8$. At $\xi = 2.81$, a sharp decrease in the radiation flux is observed. This sharp transition, as in the case of disk efficiency, is associated with a jump in the choice of $r_{ms}$.

Fig. 4b demonstrates the variation of temperature $Log[T(r)]$ in Kelvin from the inner boundary of the disk $r_{ms}$ outwards with increasing orders of tuning parameter $\xi$. The qualitative features of the temperature profiles are determined by radiation flux $F(r)$ because of the Stefan-Boltzmann law, Eq. (12). As revealed from Fig. 4b, the SABH profile of temperature is always higher than those of SBH at $\xi \leq 2.8$ and lower than SBH at $\xi > 2.8$. The maxima of temperature $T_{\max }^{SABH}(r)$ occurring at a radius shifting the inner boundary of the disk with increase of $\xi $ from $0$ to $2.8$, but when $\xi > 2.8$ the maxima of temperature $T_{\max }^{SABH}(r)$ shifting further the inner boundary of the disk. However, as seen from Table 2, the order of magnitude of the ratio $T_{\max }^{SABH}(r)/T_{\max }^{SBH}(r)\sim 9.3$ at $\xi = 2.8$ (SABH) and $\xi =0$ (SBH) and $T_{\max }^{SBH}(r)/T_{\max}^{SABH}(r)\sim 3.1$ at $\xi =6$ (SABH) and $\xi =0$ (SBH). Thus the SABH is hotter than the SBH at $\xi \leq 2.8$ and the SABH is colder than the SBH at $\xi > 2.8$.

Fig. 4c shows disk luminosity spectra $Log[\nu L(\nu )]$\ plotted against the observed frequency $\nu \lbrack Hz]$. Plots show a steep rise in $\nu L(\nu )$\ to $\sim 10^{30}-10^{31} [erg.s^{-1}]$\ occuring around $\nu \sim 10^{15}-10^{16} [Hz]$. The luminosity $\nu L(\nu )\rightarrow 0$ at $\nu \rightarrow \infty $. The maxima of the spectral luminosity $\nu L(\nu)_{\max }^{SABH}$\ and $\nu L(\nu )_{\max }^{SBH}$ occurring at $10^{30}-10^{32}[erg.s^{-1}]$. However, as seen from Table 2, the order of magnitude of the ratio $\nu L(\nu )_{\max }^{SBH}/\nu L(\nu )_{\max}^{SABH}=3.61$ at $\xi =0$\ (SBH) and $\xi =6$\ (SABH) and the order of magnitude of the ratio $\nu L(\nu )_{\max }^{SABH}/\nu L(\nu )_{\max}^{SBH}=207.8$\ at $\xi =0$\ (SBH) and $\xi =2.8$\ (SABH) shows the brightness of the SBH exceeds that of the SABH at $\xi > 2.8$ and the SABH is brighter than the SBH for $\xi \leq 2.8$.

\begin{table}[h]
\caption{The exact maximum values of the time-averaged radiation flux $F_{\max}(r)$, temperature $T_{\max }(r)$, and the emission spectra $\nu L(\nu )_{\max }$ for the accretion disk}\label{tab2}
\begin{tabular*}{\textwidth}{@{\extracolsep\fill}lccc}
\hline
$\xi$ & $F_{\max }(r)$ & $T_{\max }(r)$ & $\nu L(\nu )_{\max }$ \\ 
& $[\times 10^{14}$ $erg\cdot s^{-1}\cdot cm^{-2}]$ & $[\times 10^{4}$ $K]$
& $[\times 10^{30}$ $erg\cdot s^{-1}]$ \\
\hline
$0$ $(SBH)$ & $5.9748$ & $5.6974$ & $5.5048$ \\ 
$1$ & $37.7019$ & $9.0300$ & $19.6279$ \\ 
$2$ & $420.8985$ & $16.5060$ & $94.3466$ \\ 
$2.5$ & $6164.93$ & $34.7358$ & $486.866$ \\ 
$2.6$ & $12271.3$ & $41.6686$ & $635.278$ \\ 
$2.7$ & $19686.7$ & $47.4458$ & $846.611$ \\ 
$2.8$ & $28874.6$ & $52.9515$ & $1143.93$ \\ 
$2.81$ & $3.60835$ & $5.0225$ & $8.2546$ \\ 
$2.82$ & $3.4543$ & $4.9680$ & $8.0853$ \\ 
$2.85$ & $3.0779$ & $4.8268$ & $7.6634$ \\ 
$2.9$ & $2.6113$ & $4.6325$ & $7.1107$ \\ 
$3$ & $1.9833$ & $4.3246$ & $6.2869$ \\ 
$4$ & $0.3793$ & $2.8598$ & $3.0936$ \\ 
$5$ & $0.1384$ & $2.2229$ & $2.0504$ \\ 
$6$ & $0.0655$ & $1.8437$ & $1.5243$ \\
\hline
\end{tabular*}
\end{table}

\begin{figure}[!ht]
  \centerline{\includegraphics[scale=0.52]{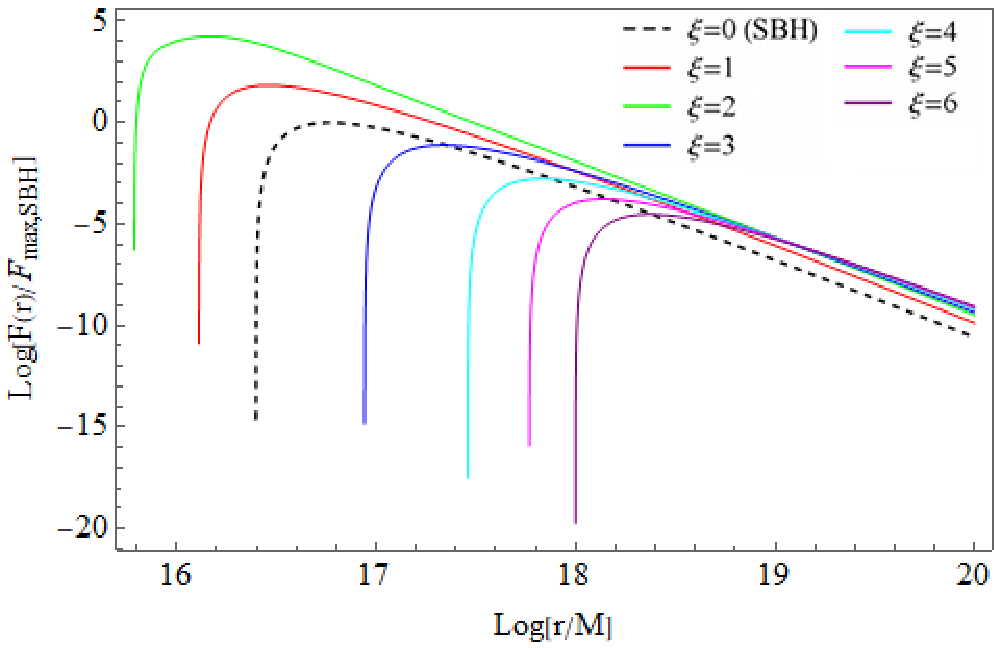} \includegraphics[scale=0.52]{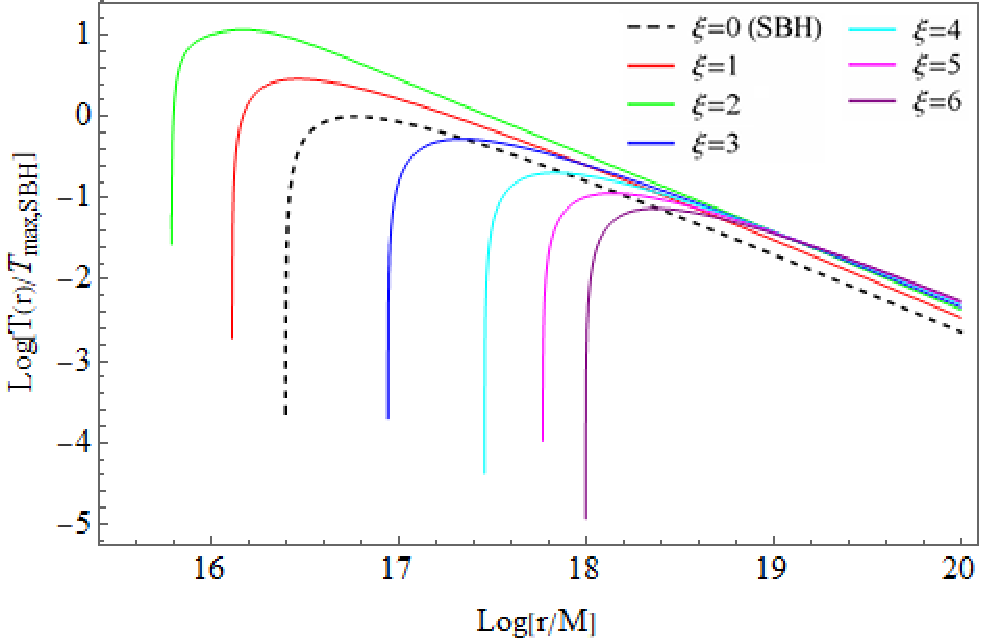}}
  \centerline{\includegraphics[scale=0.52]{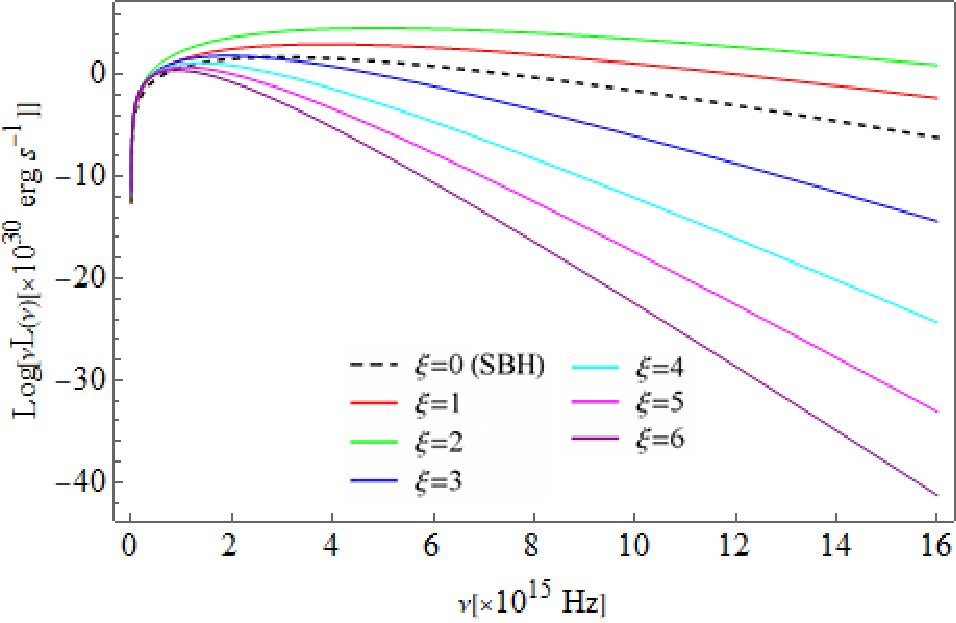}}
  \caption{The time-averaged radiation flux $F(r)$ (a), temperature distribution $T(r)$ (b), and spectral luminosity $\nu L(\nu )$ (c) of the accretion disk, for inclination $i=0$.}
\end{figure}

Thus, we have shown that for $\xi \in (0, 2.8]$, the marginally stable orbit radius decreases with increasing tuning parameter. In this case, the accretion disk becomes closer to the black hole, leading to an increase in flux, temperature, and radiation. This means that the emissivity properties of the SABH accretion disk are in a strengthening regime. However, at $\xi \in (2.8, \infty)$ a decrease in the emissivity characteristics of disk is observed, which means that the SABH operates in weakening regime.

\section{The Eddington luminosity}
\label{sec4}

Observations of Type I X-ray bursts support the theory that a star's luminosity cannot exceed the Eddington limit -- the point where radiation pressure counteracts gravity, holding the star together \cite{Frank:2002, Jonker:2004}. The Eddington luminosity at infinity \cite{Torres:2002} is given as
\begin{equation}
L_{Edd}^{\infty} = \frac{4\pi m_{p} r_{h}^{2}}{\sigma _{T}}\left( \sqrt{-\frac{g_{tt}}{g_{rr}}} \frac{d}{dr}\sqrt{-g_{tt}}\right).
\end{equation}

Substituting $r_{h}=2M$ and Eqs.(1)-(2) in Eq.(16) we obtain the Eddington luminosity at infinity of acoustic black hole as
\begin{equation}
L_{Edd}^{\infty} = 5.52 \times 10^{38} \frac{M^{3}}{r^{2}} \sqrt{\left(1 - \frac{2M}{r} \right) \left\{1 - \frac{2M\xi}{r} \left(1 - \frac{2M}{r} \right) \right\}} \left\{1 + \xi \left(1 - \frac{8M}{r} + \frac{12M^{2}}{r^{2}} \right) \right\}.
\end{equation}

Since Eq. (17) is positive only for $\xi < 3$, we define a critical radius $r_{\textmd{crit}}^{\pm}$ in which $L_{Edd}^{\infty} = 0$:
\begin{equation}
r_{\textmd{crit}}^{\pm} = \frac{2M \left(2\xi  \pm \sqrt{\xi^2 - 3 \xi}\right)}{1+\xi}.
\end{equation}

The occurrence of negative values in the formal expression for $L_{\rm Edd}^{\infty}$ should not be interpreted as a negative physical luminosity. It indicates that the static-frame gravitational acceleration changes sign. For $g_{tt}=-f(r)$ and $g_{rr}=f^{-1}(r)$, the radial acceleration of a static observer is $a^r=f'(r)/2$. Hence the Eddington balance is physically meaningful only in regions where the spacetime is static and $f'(r)>0$, so that an outward radiation force can balance an inward effective gravitational pull. In the SABH case the sign is governed by $B(r,\xi)=1+\xi(1-8M/r+12M^2/r^2)$. For $\xi>3$, $B(r,\xi)$ vanishes at $r_{\rm crit}^{\pm}$, and the interval between these roots is excluded from the usual Eddington interpretation.

\begin{figure}[!ht]
  \centerline{\includegraphics[scale=0.22]{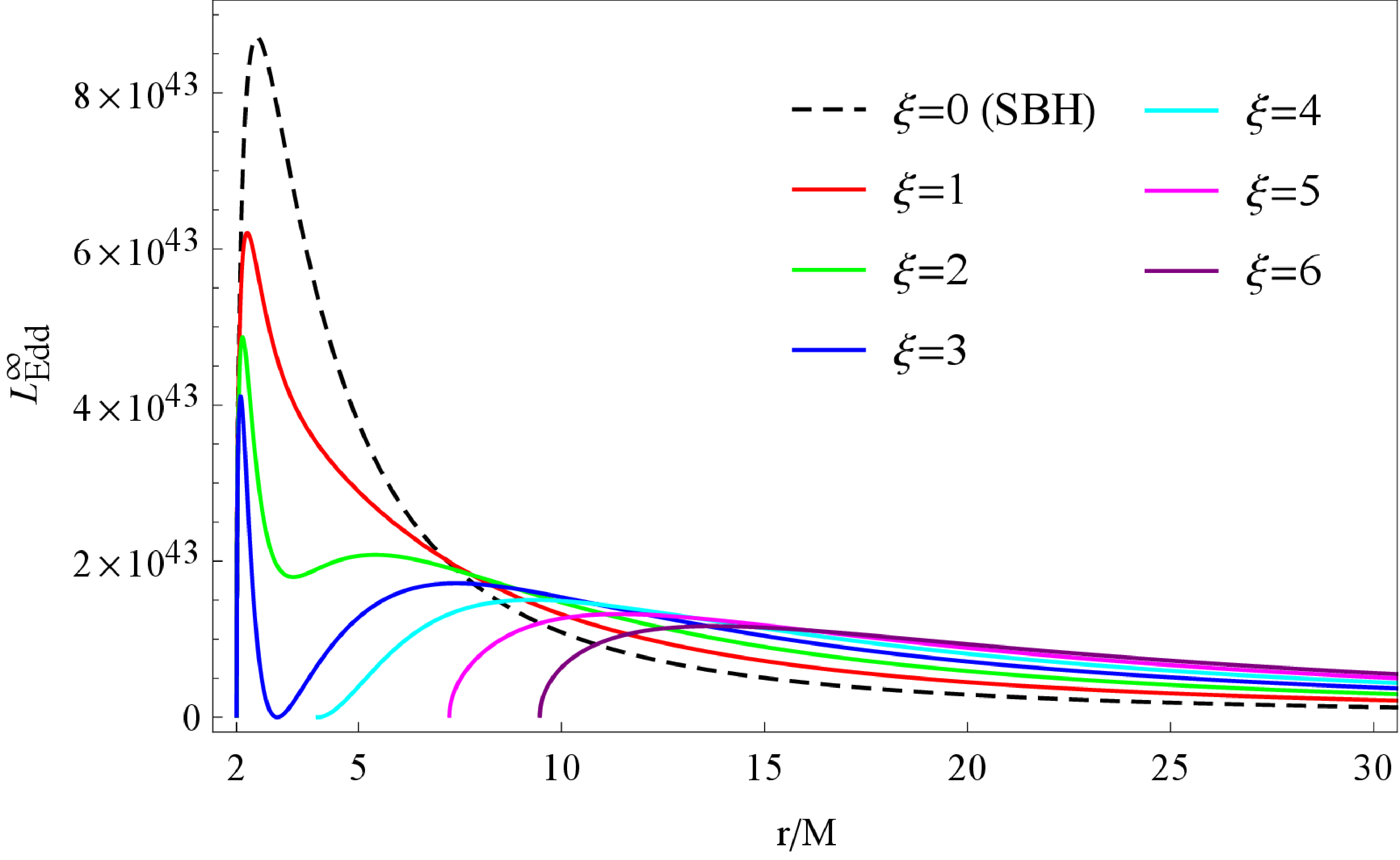}}
  \caption{The Eddington luminosity of SABH and SBH as a function of the radial coordinate $r/M$.}
\end{figure}

Fig. 5 shows the Eddington luminosity $L_{Edd}^{\infty}$ for the SABH and compared with the SBH. For values of $\xi \leq 3$, we plotted the critical luminosity in the range from $r_{h}$ to infinity, for values of $3<\xi<4$, we plotted it in the range from $r_{\min} = r_{\textmd{crit}}^{+}$ to infinity, and for values of $\xi \geq 4$, we plotted it in the range from $r_{\min} = r_{\rm ac+}$ to infinity. Thus, the critical luminosity will always be positive. As $\xi$ increases, the maximum value of luminosity $L_{Edd}^{\infty}$ of SABH decreases and is always less than SBH. Thus, here we can see that the SABH works only in the weakening regime, in contrast to the radiative properties of the thin accretion disk.

\section{Conclusion}
\label{sec5}

Studying the theoretical features of accretion disks within the framework of analogue gravity provides a unique pathway to explore the intersection of fluid dynamics and general relativity in a laboratory setting. This approach is particularly important because it allows for the simulation of strong-field gravitational effects, such as the behaviour of matter near an event horizon, using effective metrics $g_{\mu\nu}$ where the speed of sound $c_s$ plays a role analogous to the speed of light within the phenomenological interpretation adopted here. By analyzing these systems, we can gain insights into the stability and radiation signatures of astrophysical disks, as demonstrated in \cite{Sk:2025}, which examines how external environments influence emergent gravitational properties.

In this work, using the Novikov-Thorne formalism, we presented a comparative study of accretion features between SABH and SBH. This model is extremely useful and widely employed to study the accretion process. The spacetime of the SABH differs from that of the SBH in many important ways, which are expected to show up in their kinematic and emissivity properties of the thin accretion disk.

The illuminating results of the present work are the quantitative predictions of the emissivity profile maxima presented in Table 1 and Table 2. As shown there, the SABH accretion disk emission profiles can be lower than those of SBH for $\xi >2.8$, i.e., the black hole with the tuning parameter is cooler and dimmer, which means that the black hole is in the weakening regime. In particular, the highest value of the SABH flux for $\xi =6$ is $\sim 100$ times smaller than that of SBH, and the highest value of the SABH luminosity spectra for $\xi =6$ is $\sim 3.65$ times smaller than that of SBH. Thus, the greatest influence of the $\xi$ parameter can be seen in the study of the emission flux. For $\xi <2.8$, the SABH accretion disk emission profiles can be hotter and brighter than those of SBH, which indicates that black hole operates in a strengthening regime. As shown in Table 2, the maximum luminosity of SABH at $\xi = 2.8$ is $207.8$ times greater than that of SBH. The conversion efficiency of the SABH $\epsilon_{SABH}$ reaches $\sim 60.9\%$ at $\xi =2.8$.

The study also examined the Eddington luminosity of an SABH and compared the results with those of a SBH. It was shown that the highest luminosity is achieved at $\xi = 0$, i.e., for a SBH case. Increasing $\xi$ leads to a decrease in the maximum luminosity, indicating that the black hole works only in a weakening regime when studying the luminosity.

We have used $M=15M_\odot$ and $\dot{M}_{0}=10^{19}\,\mathrm{g\,s^{-1}}$ only as a toy model to illustrate the dependence of the Novikov-Thorne diagnostics on $\xi$. A direct numerical comparison with a real objects is not meaningful without rescaling the mass, accretion rate, inclination, and observing frequency. We therefore regard the Event Horizon Telescope (EHT) measurements \cite{Akiyama:2022b, Akiyama:2022c} as observational motivation for studying disk diagnostics in non-Schwarzschild geometries, rather than as data to be compared directly with the numerical values quoted here. Therefore, several issues require further study, including whether low-luminosity accretion disks exist, how jets affect luminosity, and whether the Novikov-Thorne theory can be modified to account for jets \cite{Guo:2025}.

\end{document}